\documentclass[final]{aipproc}

\usepackage{aas_macros}

\layoutstyle{6x9}

\begin{document}

\title{Non-Recycled Pulsars in Globular Clusters}


\author{Ryan S.\ Lynch} {address={Department of Astronomy, University
    of Virginia, P.O.\ Box 400325, Charlottesville, VA 22904}}

\author{Jason R.\ Boyles} {address={Department of Physics, West
    Virginia University, P.O.\ Box 6315, Morgantown, WV 26506} }

\author{Duncan R.\ Lorimer} {address={Department of Physics, West
    Virginia University, P.O.\ Box 6315, Morgantown, WV 26506} }

\author{Robert Mnatsakanov} {address={Department of Statistics, West
    Virginia University, P.O.\ Box 6330, Morgantown, WV 26506} }

\author{Philip J.\ Turk} {address={Department of Statistics, West
    Virginia University, P.O.\ Box 6330, Morgantown, WV 26506} }

\author{Scott M.\ Ransom}{ address={National Radio Astronomy
    Observatory, 520 Edgemont Road, Charlottesville, VA 22903} }

\begin{abstract}
  We place limits on the population of non-recycled pulsars
  originating in globular clusters through Monte Carlo simulations and
  frequentist statistical techniques.  We set upper limits on the
  birth rates of non-recycled cluster pulsars and predict how many may
  remain in the clusters, and how many may escape the cluster
  potentials and enter the field of the Galaxy.
\end{abstract}

\maketitle

\section{Introduction}

\citet{lmd96} noted the presence of long-period, relatively high
magnetic field pulsars in globular clusters (GCs) that seem more
similar to the ``normal'', non-recycled pulsars (NRPs) commonly found
in the Galactic disk; four such pulsars are currently known
\citep{lbhb93,bblg94,lmd96,cha03}.  Their presence is an enigma, as
NRPs are commonly assumed to form via core collapse of a massive star
and to live as pulsars for $\sim 10$--$100\; \mathrm{Myr}$, but no
massive stars have existed in GCs for $\sim 10\; \mathrm{Gyr}$.  One
popular explanation for the presence of NRPs in GCs is accretion
induced collapse (AIC) of a massive white dwarf \citep[][and
references therein]{ihr+08}, but AIC have not been directly observed
and the properties of neutron stars formed through this channel are
not well known.

We set out to use the sensitivity limits of dozens of GC pulsar
searches to estimate upper limits on the birth rates of NRPs in
clusters.  We make no assumptions about the processes leading to their
formation, so these results may be applied to a variety of channels.
We then use the implied birth rates to model a population of NRPs that
have escaped from clusters and entered the field of our Galaxy.

\section{NRP\MakeLowercase{s} Retained in Clusters}

Observing parameters from 54 GC searches were used to compute limiting
flux densities, which were then converted to limiting luminosities
($L_\mathrm{min}$) using published GC distances.  As the known NRPs in
clusters appear similar to normal Galactic pulsars, we used the
luminosity function given in \citet{f-g+k06} and a typical lifetime of
$40\; \mathrm{Myr}$.  Since GCs have relatively low escape velocities
($\sim 50\; \mathrm{km\, s^{-1}}$), many NRPs will escape the clusters
due to birth kicks.  We simulated birth rates for three Maxwellian
kick distributions with $\sigma = 10$, $130$, and $265\; \mathrm{km\,
  s^{-1}}$.  Kick velocities were chosen at random for each
distribution and compared to published escape velocities for each
cluster---if $v_{\mathrm{psr}} < v_{\mathrm{esc}}$, the pulsar was
assumed to be retained by the cluster and assigned a luminosity, also
chosen at random.  The number of trials to obtain one observable
pulsar was computed, which was taken to be an estimated upper limit
for the number of NRPs in a particular cluster for the given velocity
distribution, which was then divided by the assumed lifetime to arrive
at a birth rate.

We also used an approach that assumes the number of observable pulsars
is described by a binomial distribution.  In clusters with no known
NRPs, this reduces to $P(N,n=0) = (1 - p)^N$, where $N$ is the total
number of NRPs, $n$ is the number detected, and $p$ is the probability
that $L_{\mathrm{psr}} > L_{\mathrm{min}}$.  The results for both
approaches are similar and predict an upper limit of $\sim 5000$ NRPs
across the clusters we have studied.

\section{NRP\MakeLowercase{s} That Escape from Clusters}

The birth rates calculated above for each velocity distribution were
used to estimate the number of NRPs that will escape from their host
clusters.  The motion of pulsars with velocities sufficient to escape
the cluster were integrated through the Galactic potential.  Birth
properties were assigned according to the distributions in
\citet{f-g+k06} and evolved accordingly.  This population was then
subjected to model pulsar searches using the
\texttt{PSRPOP}\footnote{http://psrpop.sourceforge.net/} software
suite.  We used two approaches to simulate data for our models.  The
first used the exact birth rates described above, using the median
value for clusters with unknown birth rates.  However, this method
does not provide very useful constraints (i.e., the upper limits are
very large), so we also assigned each cluster a birth rate that was
scaled by the V-Band luminosity of the cluster, with M22 (our most
sensitively searched cluster) used as a reference.  Our reasoning here
is that the birth rates should scale with the number of stars in the
cluster, roughly given by the luminosity.  We are exploring other
scalings as well.

We find that it may be possible for future pulsar surveys using Square
Kilometer Array-like telescopes to detect a population of NRPs that
originated in GCs, if the birth rates are sufficiently high.  The
pulsars could be separated from field pulsars by their differing
kinematics.  However, this depends on the assumed birth rates, which
in turn depend on search sensitivities, so deeper searches are needed
to draw firm conclusions.

\begin{theacknowledgments}
  We gratefully acknowledge support from the National Science
  Foundation through grant AST-0907697.
\end{theacknowledgments}

\bibliographystyle{aipproc}

\bibliography{references}

\IfFileExists{\jobname.bbl}{}
 {\typeout{}
  \typeout{******************************************}
  \typeout{** Please run "bibtex \jobname" to obtain}
  \typeout{** the bibliography and then re-run LaTeX}
  \typeout{** twice to fix the references!}
  \typeout{******************************************}
  \typeout{}
 }

\end{document}